\documentclass{iopart}

\usepackage{graphicx}
\usepackage{sistyle}
\usepackage{bm}

\usepackage[usenames,dvipsnames]{xcolor}
\usepackage{hyperref}

\newcommand{\abs}[1]{\left| #1 \right|}
\newcommand{\bra}[1]{\left\langle #1 \right|}
\newcommand{\ket}[1]{\left| #1 \right\rangle}
\newcommand{\braket}[2]{\left\langle {#1%
{\left| \vphantom{#1 #2} \right.} #2} \right\rangle}
\newcommand{\vect}[1]{\bm{#1}}

\begin{document}

\title[Determining the dimensionality of orbital-angular-momentum entanglement]{Determining the dimensionality of bipartite orbital-angular-momentum entanglement using multi-sector phase masks}

\author{D.~Giovannini\(^1\), F.~M.~Miatto\(^2\), J.~Romero\(^{1,2}\), S.~M.~Barnett\(^2\), J.~P.~Woerdman\(^3\), M.~J.~Padgett\(^1\)}
\address{\(^1\) School of Physics and Astronomy, SUPA, University of Glasgow, Glasgow G12 8QQ, United Kingdom}
\address{\(^2\) Department of Physics, SUPA, University of Strathclyde, Glasgow G4 ONG, United Kingdom}
\address{\(^3\) Huygens Laboratory, Leiden University, PO Box 9504, 2300 RA Leiden, The Netherlands}
\ead{miles.padgett@glasgow.ac.uk}

\begin{abstract}
The Shannon dimensionality of orbital angular momentum (OAM) entanglement produced in spontaneous parametric down-conversion can be probed by using multi-sector phase analysers \cite{Pors:2008a}. We demonstrate a spatial light modulator-based implementation of these analysers, and use it to measure a Schmidt number of about 50.
\end{abstract}

\submitto{\NJP}
\maketitle

\section{Introduction}

The photon pairs generated by spontaneous parametric down-conversion (SPDC) in a non-linear crystal are correlated in their various properties, some of which can exhibit high-dimensional entanglement. The orbital angular momentum (OAM) of light is one such property, associated with phase structures of the form \(e^{i\ell\phi}\), where \(\ell\hbar\) is the OAM carried by each photon \cite{Allen:1992}. As \(\ell\) is an integer and is theoretically unbounded, OAM offers a natural discrete space for exploring high-dimensional entanglement \cite{Yao:2011a}.
In SPDC with a Gaussian pump (with \(\ell=0\)), the spectrum of OAM correlations between the signal and idler photons is peaked at \(\ell=0\), with tails towards high \(\abs{\ell}\) values. The width and shape of the spectrum, and therefore the number of OAM modes that constitute the two-photon entangled state, can be engineered directly by manipulating the structure of the beam pumping the crystal \cite{Torres:2003, Yao:2011, Miatto:2011} or by tuning the phase-matching conditions in SPDC \cite{Di-Lorenzo-Pires:2010}.

It is necessary to distinguish between the entanglement of the high-dimensional OAM state generated by the SPDC process, whose dimensionality is described by the Schmidt number \(K\) \cite{Law:2004}, and that detected by the measurement stage of the system \cite{Pors:2008a}, with Shannon dimensionality \(M\) \cite{Pors:2008}. \(M\) is in fact dependent upon both the generated down-conversion OAM spectrum and the finite detection capability of our analysers. The latter is expressed in terms of the number of modes \(D\) that an analyser has access to. By knowing the relation between these three quantities, we can infer \(K\) after measuring \(M\) with an analyser characterised by a known \(D\).

Determining the measured dimensionality \(M\) of tailored high-dimensional entangled states can be carried out by performing appropriate selective projective measurements \cite{Bennett:1996}. One such measurement is based on pairs of multi-sector phase plates, placed in the two arms of a down-conversion system. Each plate has \(N\) azimuthal angular sectors, each of which introduces a \(\pi\) phase shift. The number and angular width of the sectors of the phase analysers placed in each of the signal and idler arms define the superposition of OAM eigenmodes in which the two-photon state produced by down-conversion is projected. By optimising the binary phase profile of the phase analysers, it is possible to maximise the Shannon dimensionality \(D\) of the measurement apparatus, for any number of sectors \(N\) of the two plates.

By using angular phase analysers we infer the Schmidt number \(K\), characterising the effective number of azimuthal entangled OAM modes. In contrast to previous works, which used micromachined phase plates \cite{Pors:2011}, we implement \(N\)-sector angular phase analysers using spatial light modulators (SLMs) \cite{Leach:2009}. Computer-controlled SLMs provide a fast, convenient and reliable way of producing holographic phase masks with arbitrary orientations and numbers of sectors, to be used in the measurement of the Shannon dimensionality of OAM entanglement. The use of multi-sector phase masks to probe high-dimensional states, as opposed to narrow single-sector analysers \cite{Leach:2010, Dada:2011}, allows the measurements of tight angular correlations whilst maintaining high optical throughput.

\section{Theory}

\subsection{Amplitude and phase masks}

The angular measurement of a light field can be achieved by employing an angular slit. The idea behind this approach is the angular analog of a linear slit: a linear slit measures a field at one linear coordinate, with an uncertainty that depends on the width of the slit. An angular slit, on the other hand, measures a field at one angular coordinate with an uncertainty that depends on the angular width of the slit. As the angular position and orbital angular momentum form a pair of conjugate observables, the tighter the angular correlations, the larger the spread in the OAM observable \cite{Barnett:1990, Franke-Arnold:2004}. If one seeks high measurement accuracy, the angular slit has to be very narrow and, in turn, this means that much of the light is blocked, yielding the problem of lower number of counts as the measurement uncertainty decreases. A practiced solution to this problem is to employ phase masks, instead of amplitude masks.

The design of the phase mask consists of an angular step plate, which is characterised by a number of sectors \(N\) and by a set of angles that describe the position and width of each sector \cite{Pors:2011}. Each alternate sector applies a \(\pi\) phase shift. The overall action of a phase plate of this kind is therefore to flip the phase of a light field in each sector, about the centre of the plate, and leaving the phase unchanged everywhere else.
Note that the action of a phase plate does not affect the radial degree of freedom, as the design is radially invariant. This means that there is no coupling between different eigenmodes of the radial degree of freedom, which allows us to restrict ourselves to the azimuthal content of the measured state.
This effect can be described in terms of OAM. A plane wave is turned into a superposition of different OAM eigenstates. The range of OAM eigenstates of which the superposition consists depends on the number of sectors and on their relative positions and widths \cite{Pors:2011}. Such effect is analogous to the effect of an amplitude mask, without the drawback of letting less and less light through as the angular uncertainty is decreased.

\subsection{Measurement of Hilbert space dimensionality}

We probe the two-photon state produced by down-conversion
\begin{equation}
\ket{\Psi} = \sum_{\ell=-\infty}^\infty c_\ell \ket{\ell}_s \otimes \ket{-\ell}_i
\end{equation}
described in the OAM basis, where \(\ket{\ell}_s\) and \(\ket{\ell}_i\) correspond to the states of the signal and idler photons respectively. By expressing the projection state associated with a phase-mask analyser oriented at an angle \(\alpha\) as the superposition
\begin{equation}
\ket{A(\alpha)} = \sum_\ell \lambda_\ell\ket{\ell}\e^{i\ell\alpha},
\end{equation}
the coincidence probability for a pair of analysers, oriented at \(\alpha\) and \(\beta\) respectively, is given by
\begin{equation}
P(\alpha,\beta) = \abs{\braket{A(\alpha),B(\beta)}{\Psi}}^2,
\label{ModesOverlap}
\end{equation}
where \(\bra{A(\alpha),B(\beta)} = \bra{A(\alpha)}\otimes\bra{B(\beta)}\).
The coefficients \(\gamma_\ell=\abs{\braket{\ell}{A(0)}}^2=\abs{\lambda_\ell}^2\) (with \(\sum_\ell\gamma_\ell=1\)), defined by the profile of the \(N\)-sector phase masks, determine the respective OAM spectrum.
It is possible to design the arc sectors of each \(N\)-sector phase mask in a way that maximises the Shannon dimensionality \(D\) of the measured entanglement. For each \(N\) we used the optimal arrangement of sectors, based on results from \cite{Pors:2011}.
The maximum number of modes \(D\) that can in principle be measured by each of such optimal \(N\)-sector masks can then be inferred from the theoretical distribution of eigenmodes \(\gamma_\ell\) as \(D=1/\sum_\ell\gamma_\ell^2\), as shown in \cite{Pors:2011}.

In contrast, the Shannon dimensionality \(D\) of each \(N\)-sector mask used here was derived through a numerical model. The numerical model considered the distribution of the overlap between two identical \(N\)-sector masks within a two-dimensional region with a Gaussian profile, as the orientation \(\beta\) of one of the masks was rotated with respect to the other, \(\alpha\). The maximum measurable dimensionality \(D\) for each optimal \(N\)-sector mask is shown in fig.~\ref{DKMPlot1}. As already noted in \cite{Pors:2011}, it was found that, for \(N\) at least up to 14, \(D\) increases approximately linearly with \(N\). Given the coincidence probability distribution \(P(\alpha-\beta)\) obtained from the numerical model, the Shannon dimensionality \cite{Pors:2008a} can be directly calculated as
\begin{equation}
D = \frac{2\pi}{\int_0^{2\pi} P(\alpha-\beta) \, d(\alpha-\beta)},
\label{ShannonDim}
\end{equation}
where angle \(\beta\) is measured with respect to \(\alpha\). We implemented multi-sector masks with a number \(N\) of sectors between 1 and 16. The dimensionalities \(D\) obtained from the numerical model previously described and theoretical model, the latter obtained from the decomposition of each \(N\)-sector mask into eigenmodes \(\gamma_\ell\) and the sum over the first thousand terms of each corresponding infinite series expansion \cite{Pors:2011}, were found to differ by less than \(\SI{2.5}{\%}\).

\section{Experimental results}

\begin{figure}
\begin{center}
\includegraphics[width=0.7\linewidth, trim = 38mm 45mm 50mm 10mm, clip]{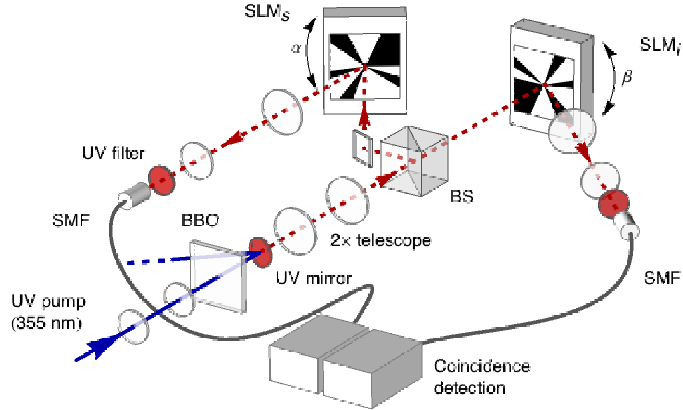}
\end{center}
\caption{\label{SPPSetup}Experimental setup implementing SLM-based coincidence detection with angular-sector phase masks (see text for details). The \(\beta\)-barium borate (BBO) crystal is pumped by a \(\SI{355}{nm}\) laser and imaged onto two spatial light modulators (SLMs). The outputs of the SLMs are then imaged onto single-mode fibres (SMFs).}
\end{figure}

\subsection{Experimental setup}

Implementing phase-mask analysers using computer-controlled spatial light modulators allows for quick and effective measurement of the Schmidt number \(K\) of the entangled state produced by SPDC. No optical elements need to be fabricated, physically rotated or replaced when using a different multi-sector mask, as the measurement process simply involves displaying one of a set of different rotated \(N\)-sector holograms on the SLMs and performing coincidence detection.

A \(\SI{5}{mm}\)-thick \(\beta\)-barium borate (BBO) non-linear crystal cut for type-I collinear SPDC acts as our source of entangled photon pairs. The crystal is pumped by a \(\SI{1}{W}\) UV laser to produce frequency-degenerate entangled photon pairs at \(\SI{710}{nm}\). The co-propagating signal and idler photons are separated by a non-polarizing beam splitter, and redirected to SLMs. The SLMs, onto which the crystal output face is imaged by a \(2\times\) telescope, are encoded with \(N\)-sector phase holograms. The SLMs are then imaged onto single-mode fibres (SMFs), which couple the SLM output to single-photon photo-diode detectors (fig.~\ref{SPPSetup}), whose output is routed to coincidence-counting electronics. The coincidence counting has a timing window of \(\SI{10}{ns}\). Narrow-band, \(\SI{2}{nm}\) interference filters are placed in front of the detectors to ensure that the frequency spread of the detected down-converted fields is small compared to the central frequencies. SLMs introduce great flexibility in our measurements, but this comes at the price of an overall lower detection efficiency, as the diffraction efficiency of SLMs is around \(\SI{50}{\%}\).

\begin{figure}
\begin{center}
\includegraphics[width=0.7\linewidth]{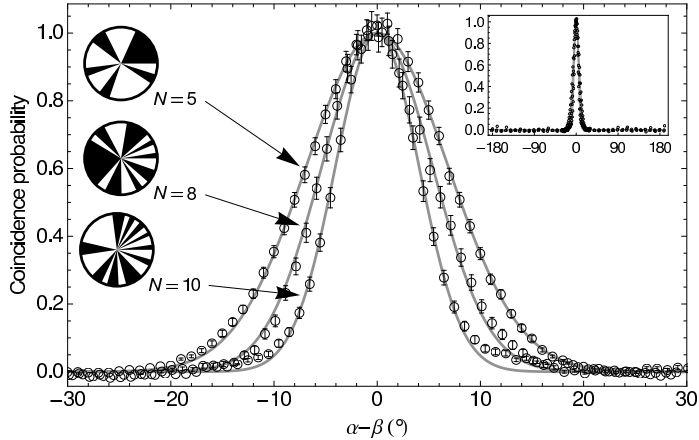}
\end{center}
\caption{\label{SampleFits}Typical best Gaussian fits of coincidence probability distributions \(P(\alpha-\beta)\) are shown for \(N=5,8,10\) and \(\phi=0\). The inset shows the full \(\ang{-180}<\alpha-\beta<\ang{180}\) range used in the measurements, for the aforementioned values of \(N\). Background subtraction was performed by assessing the experimental accidental coincidences.}
\end{figure}

This detection configuration is insensitive to any overall phase factors. Therefore, while the conservation of OAM in the SPDC process would require placing mutually phase-conjugate \(N\)-sector phase masks in the detection arms (i.e., the 0 and \(\pi\) phase-shifted sectors are inverted between the two masks), two identical phase masks can be used instead.
These phase masks are self-conjugate in case of a \(\pi\) phase shift.
The finite pixel size of the SLMs places a restriction on the width of the sectors that can be displayed on the holograms. We show that we can implement optimal multi-sector phase masks with \(N=1,\dots 16\). Suppressing the centres of the holograms, where the \(N\) angular sectors meet in a very limited spatial region of the SLM displays, did not turn out to be necessary.

The phase-matching conditions of the down-conversion process for the BBO crystal were adjusted by slightly changing the orientation of the crystal with respect to the propagation direction of the pump beam \cite{Romero:2012}. This allowed us to increase the width of the orbital angular momentum spectrum, and thus decrease the width of the angular correlations distribution. The intensity profile of the down-conversion emission can be expressed as: \(I \propto \abs{E}^2 \propto \text{sinc}(\phi+c\xi)^2\), where \(\xi\) is the external emission angle in air, \(c=(\abs{\vect{k}_s}+\abs{\vect{k}_i})L/(2n)^2\) is a constant depending on the experimental parameters, and \(\phi=(\abs{\vect{k}_p}-\abs{\vect{k}_s}-\abs{\vect{k}_i})L/2\) (with \(L\) length of the crystal) determines the degree of non-collinearity of the process \cite{Pors:2011a}. Measurements were performed for collinear (\(\phi=0\)) and near-collinear (\(\phi=-2.3\)) phase-matching conditions.

The coincidence probability distribution \(P(\alpha-\beta)\) was obtained by changing the orientation \(\beta\) of the second phase mask over the range \(\alpha\pm\ang{180}\), where \(\alpha\) is the orientation of the first. We found that, for the purposes of the experiment, a Gaussian distribution is an excellent empirical fit for the coincidence probability distributions (fig.~\ref{SampleFits}). The detected number of modes \(M\), dependent on both the source and the detectors' properties, was obtained by substituting the Gaussian fit of the measured coincidences to \(P(\alpha-\beta)\) in eq.~\eref{ShannonDim}.

\begin{figure}
\begin{center}
\includegraphics[width=0.7\linewidth]{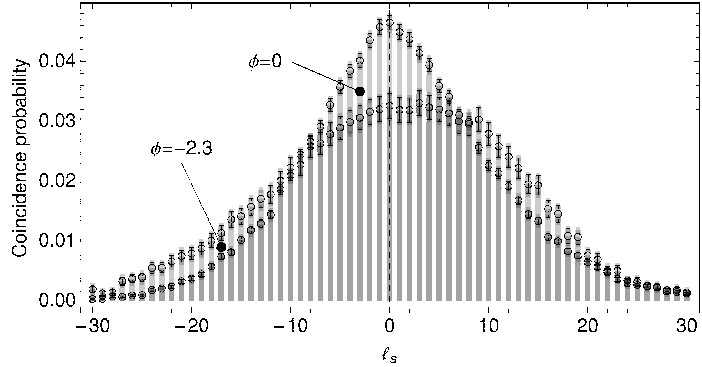}
\end{center}
\caption{\label{SBPlots} Detected orbital angular momentum spectrum with projective measurements using traditional forked holograms. Shown is the coincidence probability \(P(\ell_i, \ell_i=-\ell_s)\) for \(\ell_s\in(-30,30)\), for collinear (\(\phi=0\)) and near-collinear (\(\phi=-2.3\)) phase-matching conditions.}
\end{figure}

\subsection{Results and discussion}

The phase masks used have no radial structure. Any sensitivity of the detection apparatus to the radial quantum number \(p\) is therefore due to the spatial selectivity of the SMF coupling. Unlike full projective measurements of the down-conversion entangled state over a range of OAM eigenstates \(\ket{\ell}\) (fig.~\ref{SBPlots}), a phase-mask determination of the dimensionality \(K\) of the source does not provide any direct information on the shape of the OAM spectrum.

\begin{figure}
\begin{center}
\includegraphics[width=0.7\linewidth]{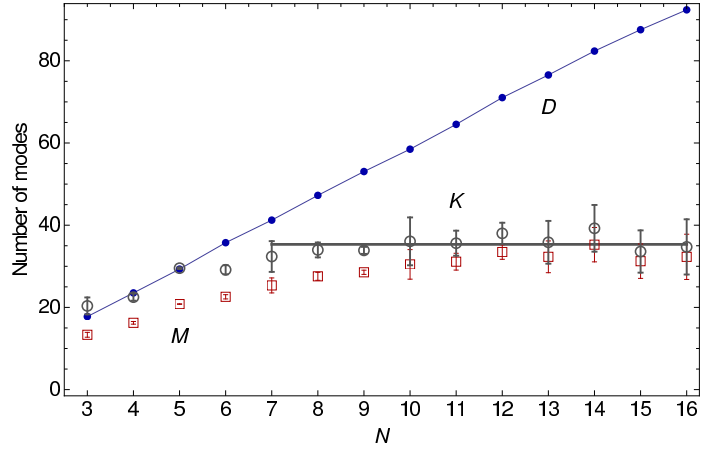}
\end{center}
\caption{\label{DKMPlot1} \emph{Collinear phase-matching conditions.} Phase mask dimensionality \(D\) from numerical model (blue points), measured \(M\) (red squares), and calculated dimensionality \(K\) (grey circles). We observe that \(D<K\) for \(N<7\); therefore, the calculated \(M\) saturates to \(D\) for any given \(N<7\). The solid grey line shows the best estimation for the number of modes \(K\) of the source, \(35 \pm 2\).}
\end{figure}

\begin{figure}
\begin{center}
\includegraphics[width=0.7\linewidth]{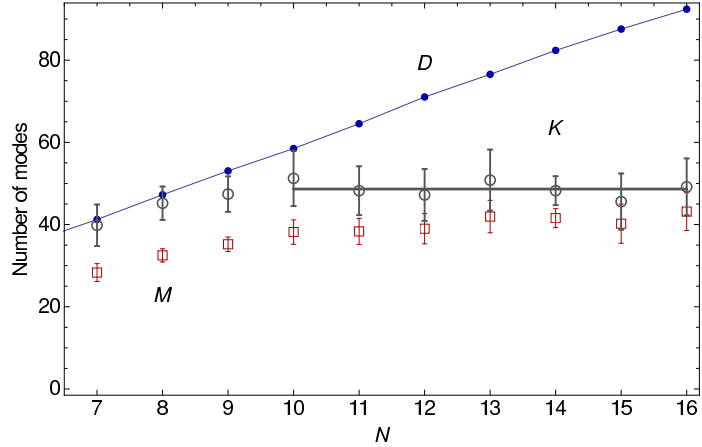}
\end{center}
\caption{\label{DKMPlot2} \emph{Near-collinear phase-matching conditions.} See caption of fig.~\ref{DKMPlot1}. Given the wider spiral bandwidth in the near-collinear regime, \(K\) saturates to \(D\) for \(N<9\). The solid grey line shows the best estimation for the number of modes \(K\) of the source, \(49 \pm 2\).}
\end{figure}

We consider the OAM spectrum generated by the source and the coincidence probability obtained from the numerical model, both fitted with Gaussian distributions, from which the dimensionalities \(K\) and \(D\) can respectively be obtained. We performed projective measurements of idler and signal over \(\ket{\ell}\) and \(\ket{-\ell}\) respectively to verify the validity of the assumption concerning the OAM spectrum (fig.~\ref{SBPlots}). The numerical model, from which the values of \(D\) for different \(N\) are obtained, calculates the overlap in eq.~\eref{ModesOverlap}.
Although the theoretical shape of the distribution of the coefficients \(\gamma_\ell\), which characterise the action of a phase mask on a light field, has own distinctive features, its implementation on an SLM makes it possible to approximate it with a Gaussian distribution.
In fact, as the hologram representing the phase plate is rotated on the surface of the SLM, any imperfections (finite pixel size, surface roughness, slight unevenness of the phase shift, electrical fluctuations) influence the effect of the phase plate in a stochastic fashion. As this a very small overall effect, it does not change the dimensionality of the phase plate but rather smooths out the distribution of the \(\gamma_\ell\) coefficients.

When the dimensionality \(K\) of the generated state is very different from the dimensionality \(D\) accessible to the detection system, the resulting measured dimensionality, \(M\), is given by the smaller of the two. For cases where \(K\) and \(D\) are comparable, \(M\) can be approximated by noting that all three distributions are close to Gaussian in form and hence their contributing widths can be combined as:
\begin{equation}
\frac{1}{M^2} \simeq \frac{1}{D^2} + \frac{1}{K^2}
\end{equation}
which gives:
\begin{equation}
M \simeq \frac{DK}{\sqrt{D^2+K^2}}
\end{equation}
from which the Schmidt number \(K\) can be derived:
\begin{equation}
K \simeq \frac{DM}{\sqrt{D^2-M^2}}.
\end{equation}
Consequently, the source dimensionality \(K\) can be inferred from the theoretical coincidence probabilities and measured coincidence distribution. The calculated dimensionality for each \(N\)-sector phase mask is shown in fig.~\ref{DKMPlot1} (collinear phase-matching) and~\ref{DKMPlot2} (near-collinear phase-matching). We measured Schmidt numbers of \(35 \pm 2\) for collinear down-conversion, and \(49 \pm 2\) in the near-collinear case.
The results for source dimensionality obtained from the phase-mask analysis are compatible with the Schmidt number derived from projective spiral bandwidth measurement (fig.~\ref{SBPlots}), with the assumption of perfect single-mode detection.

The mean visibility achieved in the experiment, defined here as the ratio between the mean baseline of the measured coincidence probability and the peak at \(\alpha-\beta=0\), without background subtraction, is \(\SI{90}{\%}\) for collinear phase-matching, and \(\SI{92}{\%}\) for near-collinear. Systematic errors due to misalignment are found to be much larger than photon statistics uncertainties.

\section{Conclusions}

We have shown how multi-sector phase-mask analysers can be implemented using spatial light modulators, and used them to probe the effective number of modes in the high-dimensional bi-photon entangled state produced by parametric down-conversion. We used a set of several multi-sector analysers to infer the Schmidt number for different phase-matching conditions, and therefore, different widths of the OAM spectrum of the source.

\section*{Acknowledgments}

This work was supported by the UK EPSRC and the EU Future and Emerging Technologies (FET) programme (HIDEAS No.~FP7-ICT-221906). We thank Hamamatsu for their support. S.M.B.~and M.J.P.~thank the Royal Society and the Wolfson Foundation for financial support.

\section*{References}

\bibliographystyle{iopart-num}
\bibliography{SectorPhaseMasks.bib}

\end{document}